\documentclass[conference]{IEEEtran}
\usepackage{ amssymb }
\usepackage{color}

\usepackage{algorithm}
\usepackage{algpseudocode} 

\usepackage{mathtools}
\DeclareMathOperator*{\argmax}{arg\,max}
\DeclareMathOperator*{\argmin}{arg\,min}

\usepackage{amsthm}
\usepackage{amsmath}
\newtheorem{theorem}{Theorem}

\title{Multi-Library Coded Caching}

\author{\IEEEauthorblockN{Saeid Sahraei\IEEEauthorrefmark{1},
Michael Gastpar\IEEEauthorrefmark{2}}
\IEEEauthorblockA{School of Computer and Communication Sciences,
EPFL\\
Lausanne, Switzerland\\
Email: \IEEEauthorrefmark{1}saeid.sahraei@epfl.ch,
\IEEEauthorrefmark{2}michael.gastpar@epfl.ch}}

\begin{document}

\maketitle
\begin{abstract}
We study the problem of coded caching when the server has access to several libraries and each user makes independent requests from every library. The single-library scenario has been well studied and it has been proved that coded caching can significantly improve the delivery rate compared to uncoded caching. In this work we show that when all the libraries have the same number of files, memory-sharing is optimal and the delivery rate cannot be improved via coding across files from different libraries. In this setting, the optimal memory-sharing strategy is one that divides the cache of each user proportional to the size of the files in different libraries. As for the general case, when the number of files in different libraries are arbitrary, we propose an inner-bound based on memory-sharing and an outer-bound based on concatenation of files from different libraries.
\end{abstract}

\section{Introduction}

The peak traffic in Content Delivery Networks (CDNs) is on the rise due to a growing demand from the users as well as an increase in the number of companies who offer streaming services. This is in contrast with the fact that during several hours per day the data traffic is relatively low and the network resources are not exploited at their full potentials. Coded caching is a strategy proposed in \cite{maddah2014fundamental} which alleviates the local memories of the users in order to decrease and smoothen the variability of network traffic over time. During the low traffic period certain globally optimized functions of the files located at the server are transferred and stored at the local caches of the users. This content placement phase will help to decrease the peak traffic of the network during the congestion period.

In this work we study the performance of a CDN which has access to the data from several different companies, as one naturally expects in practice. 
The model studied in \cite{maddah2014fundamental} considers a single collection which consists of all the files required by the users. These files are of equal size and each user is interested in precisely one such file. In this sense, the model does not distinguish among heterogeneous data, and does not take into account independent requests that a user may make from different providers. Hence, for our purpose we introduce a new model; we assume the CDN has access to multiple collections of files which we refer to as {\it libraries}. The files on different libraries are not necessarily of equal size and each user makes independent requests from different libraries.
Subsequently, our goal is to find the optimal caching strategy for such a network. That is, we are interested in minimizing the total delivery rate $R$ assuming each user has a cache of size $M$.

Our main contribution is to derive inner and outer bounds for the delivery rate of the described network, and to show that under certain constraints, the optimal caching strategy only requires coding across files within the same library. In other words, each user partitions her cache into several segments and dedicates one segment to each library and ignores coding opportunities across files from different libraries. The size of each segment should be chosen proportional to the size of the files in the corresponding library. 

The optimality of this memory-sharing strategy has interesting practical implications. Firstly, if one knows the optimal caching strategy for the single-library problem, one can simply extend it to multiple libraries. Secondly, although CDNs receive their data from multiple different companies, big software corporations such as Amazon and Netflix are moving their traffic to their own CDNs and perform independently from one another. This can be modeled as a network with several servers each having access to distinct files and having limited or no interactions among themselves. The optimality of memory-sharing implies that there is no loss due to this emigration from one centralized CDN to multiple isolated ones. From another perspective, coding across files from different servers in the placement phase introduces vulnerability to network failures; if one server goes down in the delivery phase, the users will not be able to recover the files from any other.

The basic coded caching strategy proposed in \cite{maddah2014fundamental}   has been extended to a variety of other networking scenarios, among which are decentralized \cite{maddah2013decentralized}, multi-server \cite{shariatpanahi2015multi}, hierarchical \cite{karamchandani2014hierarchical}, multi-request \cite{ji2015caching} and online coded caching \cite{pedarsani2014online}, and caching with heterogenous cache sizes \cite{wang2015fundamental}. 
Perhaps the most relevant to our work is ``multi-level coded caching" \cite{hachem2014content,hachem2014multi}     where several popularity classes of files are served to the users via access points that are in possession of local caches. The mathematical model in \cite{hachem2014multi} is similar to the model considered here,
with ``popularity classes'' taking the role of ``libraries'' in our terminology.
More precisely, the model in \cite{hachem2014multi} is slightly more general in that it allows multiple users to have access to each cache, and is slightly less general in that it forces files on all libraries to be of the same size. The more important difference between \cite{hachem2014multi} and the present paper concerns the results: The caching strategies are substantially different, and while \cite{hachem2014multi} establishes order-optimality (under specific constraints), the present paper establishes an exact optimality result (under certain other constraints).

We continue this paper by providing a motivating example in Section \ref{sec:example}. Next, we will formally define our problem in Section \ref{sec:formal} and express our main achievability and converse results in Section \ref{sec:main}. In Section \ref{sec:optimal} we will find the optimal memory-sharing strategy and will prove that it is globally optimal when the number of files are equal in different libraries.

\section{Motivating Example}
\label{sec:example}
Assume we have two libraries. Library $1$ consists of two files $A$ and $B$ each of size $F_1$ and in Library $2$ there are two other files $C$ and $D$ each of size $F_2 = 1.5F_1$. Suppose we have two users each with a cache of size $ M = F_1 + F_2$. We plan to perform memory-sharing, that is to divide the cache of each user into two segments and assign each segment to one library. Let us assume we assign $\lambda M$ of each cache to Library $1$ and the rest to the Library 2, for some $0\le \lambda \le 1$.  In the delivery phase, each user will request one file from each library. In other words, each user will request either of $\{A,C\}$, $\{A,D\}$, $\{B,C\}$ or $\{B,D\}$.

\begin{figure}
\includegraphics[scale = 0.15]{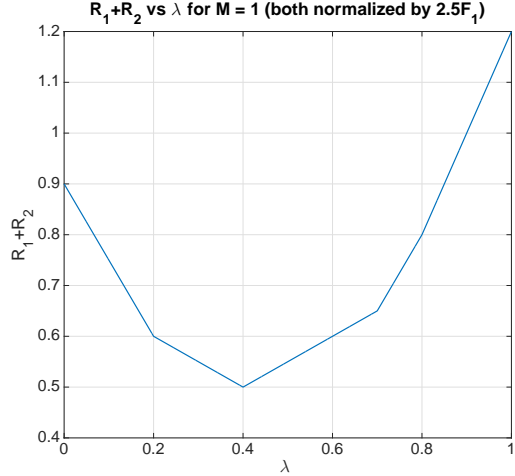}
\caption{The sum of the delivery rates of Library 1 and Library 2 vs $\lambda$, the fraction of the cache dedicated to Library $1$. Library $1$ consists of $2$ files each of size $F_1$ and Library $2$ has two files each of size $F_2 = 1.5 F_1$. The cache size of each of the two users is $M = F_1 + F_2$.}
\label{fig:example22}
\end{figure}

For each library we know the optimal coded caching strategy from \cite{maddah2014fundamental}. Therefore, for each $\lambda$ we know the minimum value of $R(\lambda) \stackrel{\triangle}{=} R_1(\lambda M) + R_2((1-\lambda)M)$. This curve is plotted in Figure \ref{fig:example22} and can be described by the following set of equations
\begin{equation*}
R(\lambda ) = \begin{cases} \frac{9}{10} - \frac{3}{2}\lambda & \mbox{ if } 0\le\lambda<\frac{1}{5}, \\
\frac{7}{10} - \frac{1}{2}\lambda  & \mbox{ if } \frac{1}{5}\le\lambda <\frac{2}{5},\\
\frac{3}{10} + \frac{1}{2}\lambda  & \mbox{ if } \frac{2}{5}\le\lambda <\frac{7}{10},\\
-\frac{2}{5} + \frac{3}{2}\lambda  & \mbox{ if } \frac{7}{10}\le\lambda <\frac{4}{5},  \\
-\frac{4}{5} + 2\lambda & \mbox{ if } \frac{4}{5}\le\lambda \le 1 .
\end{cases}
\end{equation*}
Evidently, the minimum of $R_1 + R_2$ is $\frac{1}{2}$ and is attained for $\lambda = \frac{2}{5}$. This is the same point that we obtain if we divide the cache size proportional to the size of the files on the two libraries, i.e. $\lambda = \frac{F_1}{F_1 + F_2}$. As we will see in Section \ref{sec:equalNum}, this is always the case, regardless of the number of libraries, the number of users or the size of the cache. As long as all libraries have the same number of files, the optimal memory-sharing strategy is one that divides the cache among different libraries proportional to their size of the files. More importantly, we will see that this strategy is globally optimal. That is, coding across files from different libraries cannot help in reducing the total delivery rate.

\section{Statement of the Problem}
\label{sec:formal}
Our problem statement closely follows that of \cite{maddah2014fundamental} with the difference that we classify the files into several libraries; allow the size of the files on different libraries to be different, and allow the users to request files from every library. More formally,
suppose we have $L$ libraries each consisting of $N_\ell$ files, $\ell\in[L]$.  We denote by $W^{(\ell)}_{n}$ the $n$'th file on the $\ell$'th library and assume all the files are independent. 
Furthermore, we assume the $n$'th file on the $\ell$'th library is of size $F^{(\ell)}_{n} = \alpha^{(\ell)}_{n} F$ where $\sum_{\ell = 1}^L\frac{1}{N_\ell}\sum_{n = 1}^{N_\ell}\alpha^{(\ell)}_{n} = 1$. We impose this last normalization constraint in order to make our definition of rate and cache size compatible with the single-library case with equal file sizes. It is important to note that in this paper we are mostly interested in the the scenario where the size of the files within each library are equal. In other words, $F^{(\ell)}_{n} = F^{(\ell)}$ for $n\in [N_\ell]$ and $\ell\in[L]$. This more general notation is introduced in order to facilitate the statement of our converse results in their full generality. We have $K$ users each with a cache of normalized size $M$. The caching scheme consists of two phases, the placement phase and the delivery phase. In the placement phase each user has access to all the files and stores an arbitrary function of them of size $MF$ in her cache $Z_k$. Following the notations in \cite{maddah2014fundamental}, we call these {\it caching functions}

\begin{equation}
\phi_k:\prod_{\ell =1}^L\prod_{n = 1}^{N_\ell}[2^{F^{(\ell)}_{n}}]\rightarrow [2^{\lfloor{FM\rfloor}}],\;\forall k\in[K].
\label{eqn:caching}
\end{equation}

Note that the requests are unknown in this phase and hence $\phi_k$ does not depend on them. In the delivery phase, every user requests exactly one file from each library \footnote{It is perhaps more realistic to assume each user orders {\it at most} one file from each library. However, since we are studying the worst case analysis, this will not change the results.}. The requests made to the $\ell$'th library are represented by a vector $d^{(\ell)}_{[K]}$ for every $\ell \in[L]$ where $d^{(\ell)}_{k}\in[N_\ell]$ for every $k\in[K]$. Based on this request vector an update message $X_{\{d^{(\ell)}_{[K]}\}_{\ell = 1}^L}$ of size $RF$ is then broadcast by the server to all the users. This update message naturally depends on the requests and the files
\begin{equation*}
X_{\{d^{(\ell)}_{[K]}\}_{\ell = 1}^L} =\psi_{\{d^{(\ell)}_{[K]}\}_{\ell = 1}^L}(\{W^{(\ell)}_{[N_\ell]}\}_{\ell = 1}^L)
\end{equation*}
where $\psi_{\{d^{(\ell)}_{[K]}\}_{\ell = 1}^L}$ are called the {\it encoding functions}
\begin{equation}
\psi_{\{d^{(\ell)}_{[K]}\}_{\ell = 1}^L}:\prod_{\ell =1}^L\prod_{n = 1}^{N_\ell}[2^{F^{(\ell)}_{n}}]\rightarrow [2^{\lfloor FR\rfloor}].
\label{eqn:encoding}
\end{equation}
Each user reconstructs her desired files as a function of the content of her cache $Z_k$ and the update message.

\begin{eqnarray*}
&&\hat{W}_{\{d^{(\ell)}_{[K]}\}_{\ell = 1}^L,k,i}{=} \mu_{\{d^{(\ell)}_{[K]}\}_{\ell = 1}^L,k,i}(X_{\{d^{(\ell)}_{[K]}\}_{\ell = 1}^L},Z_k),\\&&\forall k\in[K],i\in[L]
\end{eqnarray*}
where $\mu_{\{d^{(\ell)}_{[K]}\}_{\ell = 1}^L,k,i}$ are called the {\it decoding functions}
\begin{equation}
\mu_{\{d^{(\ell)}_{[K]}\}_{\ell = 1}^L,k,i}:[2^{\lfloor RF\rfloor}]\times[2^{\lfloor FM \rfloor}]\rightarrow [2^{F^{(i)}_{d^{(i)}_k}}].
\label{eqn:decoding}
\end{equation}

We say that a memory-rate pair $(R,M)$ is achievable for a network with parameters $(L,\{\alpha^{(\ell)}_{[N_\ell]}\}_{\ell = 1}^{L},N_{[L]})$ if there exists a caching strategy such that for any request vector $\{d^{(\ell)}_{[K]}\}_{\ell = 1}^L$ each user is able to recover all her desired files. In other words, if for any $\epsilon>0$ and $F$ large enough, there exist caching, encoding and decoding functions for which the probability of error

\begin{equation*}
\max_{\{d^{(\ell)}_{[K]}\}_{\ell = 1}^L\in\prod_{\ell = 1}^L[N_\ell]^{K}}\max_{k\in[K],i\in[L]}\mathbb{P}(\hat{W}_{\{d^{(\ell)}_{[K]}\}_{\ell = 1}^L,k,i}\neq W^{(i)}_{d^{(i)}_k})
\end{equation*}
can be upper bounded by $\epsilon$.
For a network with parameters $(L,\{\alpha^{(\ell)}_{[N_\ell]}\}_{\ell = 1}^{L},N_{[L]})$ the memory-rate tradeoff is defined as
\begin{eqnarray}
&&R^*(L,M,\{\alpha^{(\ell)}_{[N_\ell]}\}_{\ell = 1}^{L},N_{[L]}) \stackrel{\triangle}{=}\nonumber\\
&&\inf\left\{R:(M,R)\; \mbox{ is achievable} \right\}.
\label{eqn:optimal_defn}
\end{eqnarray}
Whenever the size of the files within each library are equal, that is $\alpha^{(\ell)}_n = \alpha^{(\ell)}$ for $n\in[N_\ell]$ and $\ell\in[L]$, we use the simplified notation $R^*(L,M,\{\alpha^{(\ell)}\}_{\ell = 1}^{L},N_{[L]})$ instead of $R^*(L,M,\{\alpha^{(\ell)}_{[N_\ell]}\}_{\ell = 1}^{L},N_{[L]})$. Our goal is to characterize the memory-rate tradeoff of a network with $L$ libraries in terms of the memory-rate tradeoffs of networks with single libraries. To this aim we find outer and inner-bounds for the $L$-library network and prove that the two bounds match in special cases.
\section{General Results: Achievability and Converse Bounds}
\label{sec:main}

\subsection{Achievability}
Our achievability results are based on a memory-sharing strategy. We divide the cache of each user into $L$ segments and assign one segment to each library. We ignore coding opportunities across files from different libraries. Note that as pointed out in the previous section we are only expressing our results for the scenario where $F^{(\ell)}_{n} = F^{(\ell)}$, that is the files within each library are of the same size. We have the following theorem.
\begin{theorem}
Let $R^*(L,M,\{\alpha^{(\ell)}\}_{\ell = 1}^L,N_{[L]})$ describe the memory-rate tradeoff as defined in \eqref{eqn:optimal_defn} where $\alpha^{(\ell)}_{n} = \alpha^{(\ell)}$ for $n \in[N_\ell]$ and for $\ell\in[L]$. Then, we have
\begin{equation*}
R^*(L,M,\{\alpha^{(\ell)}\}_{\ell = 1}^L,N_{[L]})\le \sum_{\ell = 1}^L{\alpha^{(\ell)}} R^*(1,\frac{M_\ell}{\alpha^{(\ell)}},1,N_\ell) 
\label{eqn:ach}
\end{equation*}
where $M_\ell$'s are arbitrary non-negative numbers satisfying $\sum_{\ell = 1}^LM_\ell = M$.
\label{thm:ach}
\end{theorem}
\iftrue
\begin{IEEEproof}
Consider Network $\cal{A}$ with parameters $(L,\{\alpha^{(\ell)}\}_{\ell = 1}^L,N_{[L]})$ and $L$ networks ${\cal B}^{(\ell)}$ with parameters $(1,1,N_\ell)$, $\ell\in[L]$. Suppose for every $\ell \in[L]$ a memory-rate pair $({\frac{FM_\ell }{F^{(\ell)}}},{R_\ell})  = (\frac{M_\ell}{\alpha^{(\ell)}},R_\ell)$ is achievable for Network ${\cal B}^{(\ell)}$ with files $W^{(\ell)}_{n}$. We will prove that $(M,\sum_{\ell = 1}^L\alpha^{(\ell)} R_\ell )$ is achievable for Network $\cal{A}$. Fix some $\epsilon>0$. By definition of achievability for Network ${\cal B}^{(\ell)}$, $\ell \in [L]$ there exist caching functions
\begin{eqnarray*}
&&\phi_k^{(\ell)}:[2^{F^{(\ell)}}]^{N_\ell}\rightarrow [2^{\lfloor F^{(\ell)} \frac{M_\ell}{\alpha^{(\ell)}}\rfloor}]\;,\; \forall k\in[K]
\end{eqnarray*}
encoding functions
\begin{eqnarray*}
&&\psi^{(\ell)}_{d^{(\ell)}_{[K]}}:[2^{F^{(\ell)}}]^{N_\ell}\rightarrow [2^{\lfloor F^{(\ell)}R_\ell\rfloor}] \;,\; \forall d^{(\ell)}_{[K]}\in[N_\ell]^K
\end{eqnarray*}
and decoding functions 

\begin{eqnarray*}
\mu^{(\ell)}_{d^{(\ell)}_{[K]},k}&:&[2^{\lfloor F^{(\ell)}R_\ell\rfloor}]\times[2^{\lfloor F^{(\ell)} \frac{M_\ell}{\alpha^{(\ell)}} \rfloor}]\rightarrow[2^{F^{(\ell)}}],\\&&\; \forall k\in[K], d^{(\ell)}_{[K]}\in[N_\ell]^K
\end{eqnarray*}
such that the estimates 
\begin{eqnarray*}
\hat{W}_{d^{(\ell)}_{[K]},k,\ell}&=& \mu^{(\ell)}_{d^{(\ell)}_{[K]},k}(X^{(\ell)}_{d^{(\ell)}_{[K]}},Z_k)
\end{eqnarray*}
satisfy 
\begin{equation*}
\max_{d^{(\ell)}_{[K]}\in[N_\ell]^{K}}\max_{k\in[K]}\mathbb{P}(\hat{W}_{d^{(\ell)}_{[K]},k,\ell}\neq W^{(\ell)}_{d^{(\ell)}_k}) <\epsilon
\end{equation*}
for $F = \frac{F^{(\ell)}}{\alpha^{(\ell)}}$ sufficiently large.\\
Now for Network $\cal{A}$ we define the caching functions
\begin{equation*}
\phi_k(\{W^{(\ell)}_{[N_\ell]}\}_{\ell = 1}^L) \stackrel{\triangle}{=} [\phi_k^{(1)}(W^{(1)}_{[N_1]}),\dots,\phi_k^{(L)}(W^{(L)}_{[N_L]})],\;\forall k
\end{equation*}
the encoding functions
\begin{eqnarray*}
&&X_{\{d^{(\ell)}_{[K]}\}_{\ell = 1}^L} = \psi_{\{d^{(\ell)}_{[K]}\}_{\ell = 1}^L}(\{W^{(\ell)}_{[N_\ell]}\}_{\ell = 1}^L) \stackrel{\triangle}{=} \\
&&[ \psi^{(1)}_{d^{(1)}_{[K]}}(W^{(1)}_{[N_1]}), \dots,\psi^{(L)}_{d^{(L)}_{[K]}}(W^{(L)}_{[N_L]})],\;\; \forall\{d^{(\ell)}_{[K]}\}_{\ell = 1}^L
\end{eqnarray*}
and the decoding functions
\begin{eqnarray*}
&&\mu_{\{d^{(\ell)}_{[K]}\}_{\ell = 1}^L,k,i}(X_{\{d^{(\ell)}_{[K]}\}_{\ell = 1}^L},Z_k)\\
&&\stackrel{\triangle}{=} \mu^{(i)}_{d^{(i)}_{[K]},k}(\psi^{(i)}_{d^{(i)}_{[K]}}(W^{(i)}_{[N_i]}),\phi^{(i)}_k(W^{(i)}_{[N_i]})),\;\; \forall i,k, \{d^{(\ell)}_{[K]}\}_{\ell = 1}^L.
\end{eqnarray*}
The probability of error of this caching scheme is thus

\begin{eqnarray*}
&&\hspace{-35pt}\max_{\{d^{(\ell)}_{[K]}\}_{\ell = 1}^L\in\prod_{\ell = 1}^L[N_\ell]^{K}}\max_{k\in[K],i\in[L]}\mathbb{P}(\hat{W}_{\{d^{(\ell)}_{[K]}\}_{\ell = 1}^L,k,i}\neq W^{(i)}_{d^{(i)}_k})\\
&&\hspace{-35pt}=\max_{i\in[L]}\max_{d^{(i)}_{[K]}\in[N_i]^{K}}\max_{k\in[K]}\mathbb{P}(\hat{W}_{d^{(i)}_{[K]},k,i}\neq W^{(i)}_{d^{(i)}_k})\\
\le \epsilon.
\end{eqnarray*}
\normalsize

Furthermore, this scheme has a rate equal to $\frac{1}{F}\sum_{\ell = 1}^L F^{(\ell)} R_\ell = \sum_{\ell = 1}^L \alpha^{(\ell)} R_\ell $ and a memory of size $\frac{1}{F}\sum_{\ell = 1}^L\frac{F^{(\ell)} M_\ell}{\alpha^{(\ell)}} = M$. Therefore, the memory-rate pair $(M,\sum_{\ell = 1}^L\alpha^{(\ell)} R_\ell)$ is achievable for Network $\cal{A}$ which proves the theorem.

\end{IEEEproof}
\fi

\subsection{Converse}
Consider Network $\cal{A}$ with $L$ libraries. Roughly speaking, we will prove that any caching strategy for  Network $\cal{A}$ can also be used for Network $\cal{B}$ which has a single library consisting of files that are concatenation of files from different libraries of $A$. Intuitively, this is done by breaking each file on Network $\cal{B}$ into its subfiles and assuming that each subfile belongs to a separate library. This is formally stated in the next theorem.

\begin{theorem}
Let $R^*(L,M,\{\alpha^{(\ell)}\}_{\ell = 1}^L,N_{[L]})$ describe the memory-rate tradeoff as defined in \eqref{eqn:optimal_defn} where $\alpha^{(\ell)}_{n}  = \alpha^{(\ell)}$ for $n \in [N_\ell]$ and for $\ell\in[L]$. Furthermore, assume without loss of generality that $N_1\le N_2\le\dots\le N_L$. Then, we have
\begin{eqnarray*}
&&R^*(L,M,\{\alpha^{(\ell)}\}_{\ell = 1}^L,N_{[L]}) \ge\\
&& R^*(1,M,\beta_{[N_L]},N_L).
\end{eqnarray*}
The coefficients $\beta_{n}$ for $n\in[N_L]$ are defined as
\begin{equation*}
\beta_{n} = \frac{\sum_{i = f(n)}^L\alpha^{(i)}}{\sum_{\ell = 1}^L \alpha^{(\ell)}N_\ell}N_L
\end{equation*}
where $f(n)$ returns the smallest $j\in[L]$ such that $n\le N_j$.
\label{thm:converse}
\end{theorem}
\begin{IEEEproof}
Consider Network $\cal{A}$ with parameters $(L,\{\alpha^{(\ell)}\}_{\ell = 1}^L,N_{[L]})$ and Network $\cal{B}$ with parameters $(1,\beta_{[N_L]},N_L)$. Suppose a memory-rate pair $(R,M)$ is achievable for Network $\cal{A}$. We will prove that the same memory-rate pair $(R,M)$ is also achievable for Network $\cal{B}$. We represent the files on Network $\cal{B}$ by $W_{n}$ which are of size $\beta_{n}F = \frac{\sum_{i = f(n)}^L \alpha^{(i)}}{\sum_{\ell =1}^L\alpha^{(\ell)}N_\ell}FN_L$ for $n \in [N_L]$. We break each $W_{n}$ into disjoint subfiles 
\begin{equation}
W_{n} = [W^{(f(n))}_{n},W^{(f(n)+1)}_{n},\dots,W^{(L)}_{n}]
\end{equation}
where $W^{(\ell)}_{n}$ is of size $\alpha^{(\ell)}\frac{F}{\sum_{\ell = 1}^L \alpha^{(\ell)}N_\ell}N_L$. Fix some $\epsilon >0$. By definition of achievability for Network $\cal{A}$ with files $\{W^{(\ell)}_{[N_\ell]}\}_{\ell = 1}^L$ there exist caching functions $\phi_k$, encoding functions $\psi_{\{d^{(\ell)}_{[K]}\}_{\ell = 1}^L}$ and decoding functions $\mu_{\{d^{(\ell)}_{[K]}\}_{\ell = 1}^L,k,i}$ as in equations \eqref{eqn:caching},\eqref{eqn:encoding},\eqref{eqn:decoding} such that for any request vector $\{d^{(\ell)}_{[K]}\}_{\ell = 1}^L$ and for $F$ large enough, each user can recover her desired files with probability of error bounded by $\epsilon$.

Now for Network $\cal{B}$ we define the caching functions  
\begin{equation*}
\phi^{'}_k (W_{[N_L]}) \stackrel{\triangle}{=}  \phi_k(\{W^{(\ell)}_{[N_\ell]}\}_{\ell =1}^L),\; \forall k\in[K]
\end{equation*}
the encoding functions
\begin{equation*}
\psi^{'}_{d'_{[K]}}(W_{[N_L]}) \stackrel{\triangle}{=} \psi_{\{d^{(\ell)}_{[K]}\}_{\ell = 1}^L}(\{W^{(\ell)}_{[N_\ell]}\}_{\ell =1}^L),\; \forall d'_{[K]}\in[N_L]^K
\end{equation*}
and the decoding functions
\begin{eqnarray}
\label{eqn:discard}
&&\mu^{'}_{d'_{[K]},k} \stackrel{\triangle}{=} \\
&&\hspace{-30pt}[\mu_{\{d^{(\ell)}_{[K]}\}_{\ell = 1}^L,k,f(d'_k)},\mu_{\{d^{(\ell)}_{[K]}\}_{\ell = 1}^L,k,f(d'_k)+1},\dots, \mu_{\{d^{(\ell)}_{[K]}\}_{\ell = 1}^L,k,L}]\nonumber\\
&&\forall k\in[K], d'_{[K]}\in[N_L]^K\nonumber
\end{eqnarray}
where $d^{(\ell)}_k \stackrel{\triangle}{=} \min(d'_k,N_\ell)$. Note that if $N_\ell < d'_k$, then $d^{(\ell)}_k$ is a dummy request and the reconstructed $W^{(\ell)}_{d^{(\ell)}_k}$ will be discarded as visible from equation \eqref{eqn:discard}. The probability of error of this caching scheme is less than $L\epsilon$
\small
\begin{eqnarray*}
&&\hspace{-20pt}\max_{d^{'}_{[K]}\in[N_L]^{K}}\max_{k\in[K]}\mathbb{P}({\hat{W}}'_{d'_{[K]},k}\neq W_{d'_k})\\
&&\hspace{-20pt}=\max_{\stackrel{(d^{'}_{[K]})\in[N_L]^{K}}{d^{(\ell)}_k = \min(d'_k,N_\ell)}}\max_{k\in[K]}\mathbb{P}\left(\bigvee_{i =f(d'_k)}^L(\hat{W}_{\{d^{(\ell)}_{[K]}\}_{\ell = 1}^L,k,i}\neq W^{(i)}_{d^{(i)}_k})\right)\\
&&\hspace{-20pt}\le\max_{\{d^{(\ell)}_{[K]}\}_{\ell = 1}^L\in\prod_{\ell = 1}^L[N_\ell]^{K}} \max_{k\in[K]}\mathbb{P}\left(\bigvee_{i =1}^L(\hat{W}_{\{d^{(\ell)}_{[K]}\}_{\ell = 1}^L,k,i}\neq W^{(i)}_{d^{(i)}_k})\right)\\
&&\hspace{-20pt}\le L\max_{\{d^{(\ell)}_{[K]}\}_{\ell = 1}^L\in\prod_{\ell = 1}^L[N_\ell]^{K}}\max_{k\in[K],i\in[L]}\mathbb{P}\left(\hat{W}_{\{d^{(\ell)}_{[K]}\}_{\ell = 1}^L,k,i}\neq W^{(i)}_{d^{(i)}_k}\right)\\
&&\hspace{-20pt} \le L \epsilon.
\end{eqnarray*}
\normalsize
Since we are reusing the same caching, encoding and decoding functions, the memory-rate pair $(R,M)$ is the same for both Networks $\cal{A}$ and $\cal{B}$. Since this is one achievable strategy for Network $\cal{B}$, we have $R^*(1,M,\beta_{[N_L]},N_L)\le R^*(L,M,\{\alpha^{(\ell)}\}_{\ell =1}^L,N_{[L]})$.
\end{IEEEproof}

\section{Optimality Results}
\label{sec:optimal}
\subsection{Libraries with Equal Number of Files}
\label{sec:equalNum}
Suppose we have $N_\ell = N$ for $\ell \in[L]$. That is, all the libraries keep hold of equal number of files. We will show that if in Theorem \ref{thm:ach}     the $M_\ell$'s are chosen proportional to $F^{(\ell)}$, our inner and outer bounds will match. This implies that the simple memory-sharing strategy proposed in Theorem \ref{thm:ach} is indeed optimal and cannot be outperformed by coding across files from different libraries.
\begin{theorem}
Let $R^*(L,M,\{\alpha^{(\ell)}\}_{\ell = 1}^L,N_{[L]})$ describe the memory-rate tradeoff as defined in \eqref{eqn:optimal_defn} where $\alpha^{(\ell)}_{n}  = \alpha^{(\ell)}$ for $n\in [N_\ell]$ and for $\ell\in[L]$. Suppose we have $N_\ell = N$ for $\ell \in [L]$ and $M_\ell = {\alpha^{(\ell)}}M$. Then
\small
\begin{equation}
R^*(L,M,\{\alpha^{(\ell)}\}_{\ell =1}^L,N_{[L]}) = \sum_{\ell = 1}^L \alpha^{(\ell)} R^*(1,\frac{M_\ell}{\alpha^{(\ell)}},1,N_\ell).
\end{equation}
\normalsize
\end{theorem}
\begin{IEEEproof}
From Theorem \ref{thm:converse} we have
\begin{eqnarray}
&&R^*(L,M,\{\alpha^{(\ell)}\}_{\ell =1}^L,N_{[L]}) \ge \nonumber\\
&&R^*(1,M,\beta_{[N_L]},N_L) = R^*(1,M,1,N).
\label{eqn:first}
\end{eqnarray}
On the other hand, from Theorem \ref{thm:ach} we know that
\begin{eqnarray}
&&R^*(L,M,\{\alpha^{(\ell)}\}_{\ell =1}^L,N_{[L]}) \le\nonumber \\
&&\sum_{\ell = 1}^L{\alpha^{(\ell)}} R^*(1,\frac{\alpha^{(\ell)}M}{\alpha^{(\ell)}},1,N_\ell) =\nonumber \\
&&\sum_{\ell = 1}^L{\alpha^{(\ell)}} R^*(1,M,1,N) =R^*(1,M,1,N).
\label{eqn:second}
\end{eqnarray}
The claim follows from \eqref{eqn:first} and \eqref{eqn:second}. 
\end{IEEEproof}

\subsection{Libraries with Arbitrary Number of Files}
\label{sec:unequalNum}
In this section we find the optimal memory-sharing strategy when the number of files in different libraries are not necessarily equal. Whether this optimal memory-sharing strategy is globally optimal or not, is a question that we have no answer for at this point (but we conjecture that it is).

We know that the memory-rate tradeoff for a network with one library is convex. We will further assume that it is piecewise linear and has the following form
\begin{equation}
R^*(1,{M}{},1,N) = \begin{cases} \zeta_{0}^{(N)} - \gamma_{0}^{(N)} {M}{} &\mbox{ if }\;\;0\le {M}{}<\theta_{1}^{(N)}\\
\zeta_{1}^{(N)} - \gamma_{1}^{(N)} {M}{} &\mbox{ if }\;\;\theta_{1}^{(N)}\le {M}{}<\theta_{2}^{(N)}\\
\dots&\\
\zeta_{r-1}^{(N)}- \gamma_{r-1}^{(N)} {M}{} &\mbox{ if }\;\;\theta_{r-1}^{(N)}\le {M}{}<  N
\end{cases}
\label{eqn:plinear}
\end{equation}
where 
$\gamma_0^{(N)} >\dots>\gamma_{r-1}^{(N)} >0$ 
 (due to convexity) and $\zeta_{i-1}^{(N)}  - \gamma_{i-1}^{(N)} \theta_{i}^{(N)}  = \zeta_{i} ^{(N)} -\gamma_{i}^{(N)} \theta_{i}^{(N)} $ for $i\in[r]$ (due to continuity) and $r$ naturally depends on $N$ but to simplify the notation we have used $r = r_N$. Also define $\gamma^{(N)}_{-1}= \infty$, $\gamma^{(N)}_{r}= 0$, $\zeta^{(N)}_{-1}= \infty$, $\zeta^{(N)}_{r}= 0$, $\theta^{(N)}_{0}= 0$, $\theta^{(N)}_{r}  = N$ and $\theta^{(N)}_{r+1}  = \infty$.

Note that if $R^*(1,{M}{},1,N)$ is not piecewise linear, we can readily generalize our results by approximating $R^*(1,{M}{},1,N)$ with a piecewise linear function of arbitrarily large number of pieces. We can now describe the optimal memory-sharing strategy for the $L$-library setting.
\begin{theorem}
Suppose the memory-rate tradeoff for a network with parameters $(1,1,N)$ has the general form of  \eqref{eqn:plinear} with $r_N$ segments. Then there exists an optimal memory-sharing strategy for a network with parameters $(L,\{\alpha^{(\ell)}\}_{\ell =1}^L,N_{[L]})$, i.e. a solution to
\begin{equation*}
M^*_{[L]} = \argmin_{{M_{[L]},}{ \sum_{\ell = 1}^L M_\ell = M}} \sum_{\ell = 1}^L \alpha^{(\ell)} R^*(1,\frac{M_\ell}{\alpha^{(\ell)}},1,N_\ell)
\end{equation*}
that satisfies the following. There exist an $\hat{\ell}\in[L]$ and $L$ integers $0\le i_\ell\le r_{N_\ell},\;\ell\in[L]$ such that
\begin{equation*}
M^*_{\ell} = \begin{cases}\theta^{(N_\ell)}_{i_\ell}\alpha^{(\ell)} &\mbox{ if } \ell\neq \hat{\ell}\\
\theta^{(N_\ell)}_{i_\ell}\alpha^{(\ell)} + M_{\text{rem}}&\mbox{ if } \ell = \hat{\ell}.
\end{cases}
\end{equation*}
where $0\le M_{\text{rem}}< \alpha^{(\hat{\ell})}(\theta^{(N_{\hat{\ell}})}_{i_{\hat{\ell}}+1}-\theta^{(N_{\hat{\ell}})}_{i_{\hat{\ell}}})$ and

\begin{equation*}
\frac{\gamma_{i_{\ell}}^{(N_\ell)}}{\alpha^{(\ell)}}\le \frac{\gamma_{{i_{\ell'}}-1}^{(N_{\ell'})}}{\alpha^{(\ell')}}\;,\;\forall \ell,\ell'\in[L]
\end{equation*}
  and
\begin{equation}
\frac{\gamma_{{i_{\ell}}}^{(N_\ell)}}{\alpha^{(\ell)}}\le \frac{\gamma_{i_{\hat{\ell}}}^{(N_{\hat{\ell}})}}{\alpha^{(\hat{\ell})}}\;,\;\forall \ell \in[L].
\end{equation}
\label{thm:unequal}
\end{theorem}
\begin{IEEEproof}
Assume there exist $\ell,\ell'\in[L]$ such that $\ell\neq \ell'$ and $M^*_\ell  = \theta^{(N_\ell)}_{i_\ell}\alpha^{(\ell)}+ M_{\text{rem}}$ and $M^*_{\ell'}  = \theta^{(N_{\ell'})}_{i_{\ell'}}\alpha^{(\ell')} + M'_{\text{rem}}$ and $M_{\text{rem}}\neq 0$ and $M'_{\text{rem}}\neq 0$. Assume without loss of generality that $\frac{\gamma_{{i_{\ell}}}^{(N_\ell)}}{\alpha^{(\ell)}}\ge\frac{\gamma_{{i_{\ell'}}}^{(N_{\ell'})}}{\alpha^{(\ell')}}$. Now we set 
\begin{eqnarray*}
\delta &=& \min(\alpha^{({\ell})}(\theta^{(N_{{\ell}})}_{i_{{\ell}}+1}-\theta^{(N_{{\ell}})}_{i_{{\ell}}})-M_{\text{rem}},M'_{\text{rem}}),\\
M^*_\ell &\leftarrow & M^*_\ell + \delta,\\3x
M^*_{\ell'} &\leftarrow & M^*_{\ell'} - \delta.
\end{eqnarray*}
This moves either of $M^*_\ell$ or $M^*_{\ell'} $ (or both) to a corner point (that is, either of $M_{\text{rem}}$ or $M'_{\text{rem}}$ will be zero). Furthermore, this changes the total rate by
\begin{eqnarray*}
\Delta R = (\frac{\gamma_{{i_{\ell'}}}^{(N_{\ell'})}}{\alpha^{({\ell'})}} - \frac{\gamma_{{i_{\ell}}}^{(N_\ell)}}{\alpha^{({\ell})}})\delta\le 0.
\end{eqnarray*}
Therefore, there exists an optimal solution for which at most one of the libraries has $M_{\text{rem}}\neq 0$. We call this library $\hat{\ell}$. Next assume there exists a pair $\ell\neq\ell'\in[L]\backslash\{\hat{\ell}\}$ for which $\frac{\gamma_{i_{\ell}}^{(N_\ell)}}{\alpha^{(\ell)}}> \frac{\gamma_{{i_{\ell'}}-1}^{(N_{\ell'})}}{\alpha^{(\ell')}}$. This time we define $\delta  =\min(\alpha^{({\ell})}(\theta^{(N_{\ell})}_{i_{\ell}+1}-\theta^{(N_{\ell})}_{i_{\ell}}),\alpha^{(\ell')}(\theta^{(N_{\ell'})}_{i_{\ell'}}-\theta^{(N_{\ell'})}_{i_{\ell'}-1}))$. Again setting $M^*_\ell \leftarrow M^*_\ell +\delta$ and $M^*_{\ell'} \leftarrow M^*_{\ell'} - \delta$ results in $\Delta R< 0$.
Finally assume $ \frac{\gamma_{{i_{\ell}}}^{(N_\ell)}}{\alpha^{(\ell)}}> \frac{\gamma_{i_{\hat{\ell}}}^{(N_{\hat{\ell}})}}{\alpha^{(\hat{\ell})}}$ for some $\ell$. We can set $\delta =\min(\alpha^{(\ell)}(\theta^{(N_{\ell})}_{i_{\ell}+1}-\theta^{(N_{\ell})}_{i_{\ell}}),M_{\text{rem}})$ and $M^*_\ell \leftarrow M^*_\ell +\delta$ and $M^*_{\hat{\ell}} \leftarrow M^*_{\hat{\ell}} - \delta$ which results in $\Delta R< 0$ unless if $M_{\text{rem}}=0$, in which case we simply choose the library with the largest $\frac{\gamma_{{i_{\ell}}}^{(N_\ell)}}{\alpha^{(\ell)}}$ to be $\hat{\ell}$.
\end{IEEEproof}

The solution described by Theorem \ref{thm:unequal} can be found in an incremental way. Assume that initially the size of the cache is zero and we gradually increase it up to $M = \sum_{\ell = 1}^L\alpha^{(\ell)}N_\ell$. At any point we must decide how much of the cache should be allocated to each library. At the beginning it is advantageous to assign all the cache to library $\ell$ with the largest $\frac{\gamma_0^{(N_\ell)}}{\alpha^{(\ell)}}$, since this reduces the total delivery rate by the largest factor. This is the library which is called $\hat{\ell}$ in the theorem. This assignment continues until this library reaches the corner point $M_{\hat{\ell}} = \alpha^{(\hat{\ell})}\theta^{(N_{\hat{\ell}})}_1$. At this point $\hat{\ell}$ is re-initialized as the library with the largest right-slope and the process continues. This procedure is summarized in Algorithm \ref{Algorithm1}.

As a final remark, we conjecture that this memory-sharing strategy is again globally optimal and that our converse bound is tight in this general. In other words, we conjecture
\begin{equation*}
 \sum_{\ell = 1}^L \alpha^{(\ell)} R^*(1,\frac{M^*_\ell}{\alpha^{(\ell)}},1,N_\ell){=} R^*(1,M,\beta_{[N_L]},N_L).
\end{equation*}
This would imply that even in the general case, there is no gain from coding across files from different libraries and memory-sharing suffices for minimizing the delivery rate. 

\begin{algorithm}
\caption[caption]{Optimal Memory Allocation for $L$ libraries}
\begin{algorithmic}[1]
 \State Set AllocM$ = 0$.
 \State Set $M_\ell = 0$ for $\ell\in[L]$.
\State Set $i_\ell = 0$, for $\ell\in[L]$.
\Statex{}
\While{AllocM$<M$}
\State Find the library $\hat{\ell}$ that has the largest right slope \begin{equation*}\hspace{-30pt}\hat{\ell} = \argmax_{\ell\in[L]} \frac{\gamma_{i_{\ell}}^{(N_\ell)}}{\alpha^{(\ell)}}.\end{equation*}
\State Set $\delta = \frac{\gamma_{i_{\hat{\ell}}}^{(N_{\hat{\ell}})}}{{\alpha^{(\hat{\ell})}}}\left({\theta_{i_{\hat{\ell}}+1}^{(N_{\hat{\ell}})}}-{\theta_{i_{\hat{\ell}}}^{(N_{\hat{\ell}})}}\right)$.
\If {$M  \ge $AllocM +$\delta$} 
\State $M_{\hat{\ell}} = M_{\hat{\ell}} + \delta$.
\State AllocM $=$AllocM + $\delta$. 
\State $ i_{\hat{\ell}} = i_{\hat{\ell}} + 1.$
\Else
\State $M_{\hat{\ell}} = M_{\hat{\ell}} + M-$AllocM.
\State AllocM$ = M$.
\EndIf
\EndWhile
\State Output $M_\ell$ for $\ell \in[L]$.
\end{algorithmic}
\label{Algorithm1}
\end{algorithm}

\section{Conclusion}
In this work we studied the problem of coded caching when the server has access to several libraries. We proved that if the number of files in different libraries are all equal, memory-sharing is optimal and the delivery rate cannot be improved via coding across different libraries. This optimality has interesting practical implications regarding Content Delivery Networks that receive their data from several different companies. For the general scenario when the number of files in different libraries are arbitrary, we found an inner-bound based on memory-sharing and an outer-bound based on concatenation of files from different libraries. Future work will study the optimality of the proposed memory-sharing strategy and aim at proving that our inner and outer bounds will match regardless of the number of files on different libraries. 
   

\section*{Acknowledgement}
        \thanks{This work was supported in part by the European ERC Starting Grant 259530-ComCom.}

\bibliographystyle{IEEEtran}
\bibliography{IEEEfull,MultiLibraryCaching}

\end{document}